\documentclass{cup-hpl}

\usepackage[T1]{fontenc}
\usepackage{lmodern}
\usepackage{ragged2e}
\usepackage{enumitem}
\usepackage{graphicx}
\usepackage{placeins}
\usepackage{subcaption}

\makeatletter
\def\ps@arxivfirst{  \let\@oddhead\@empty
  \let\@evenhead\@empty
  \def\@oddfoot{\hfil\thepage\hfil}  \let\@evenfoot\@oddfoot
}
\makeatother

\begin{document}

\shorttitle{}
\shortauthor{}

\title{Stable, High-Charge, Laser-Driven VHEE Beams for Preclinical Applications}

\author[1]{Chaitanya Varma}
\author[1,2,*]{Camilla Giaccaglia}
\author[1]{Julien Gautier}
\author[1]{Jean-Philippe Goddet}
\author[1]{Amar Tafzi}
\author[1]{Pascal Rousseau}
\author[1]{Alessandro Flacco\corresp{\email{alessandro.flacco@ensta.fr}}}

\address[1]{Laboratoire d'Optique Appliquee (LOA), ENSTA Paris, Institut Polytechnique de Paris, 91762 Palaiseau, France}
\address[2]{Institut Curie, Inserm U 1021-CNRS UMR 3347, University Paris-Saclay, PSL Research University, Centre Universitaire, Orsay 91471, France}

\begin{abstract}
       \justifying
       Laser-driven Plasma Accelerators (LPAs) are emerging as promising candidates for the generation of very-high-energy-electron (VHEE) beams suitable for medical applications, especially when compact acceleration distance and ultrahigh instantaneous dose rates are desirable. Simultaneously increasing charge-per-pulse and maintaining sufficient pointing stability for VHEE-beams remains a key experimental challenge for LPA-applications like radiobiological studies. Here, we demonstrate the generation of laser-accelerated beams with total charge-per-pulse reaching (559 $\pm$ 56) pC using an 80~TW laser system. The acceleration process provides a dominant spectral component in the range of 40-60~MeV, accompanied by a persistent high-energy tail extending to approximately 200~MeV. Pointing jitter measured in air at the irradiation target plane reaches as low as 0.7~mrad. The beams were successfully used for VHEE irradiation experiments involving the irradiation of a 10~mm region of interest (ROI) through a 40~mm depth at 400 $\pm$ 20~mGy/pulse. Spectral reproducibility is quantified using energy distribution measurements over fixed-setting shot sequences. Modelling the experimental interaction via Particle-in-Cell (PIC) simulations show the formation of a peak-and-tail spectral structure as a consequence of electron acceleration in successive bubbles associated with the evolution of a strongly modified plasma wake. This provides a plausible mechanism for maintaining substantial accelerated charge despite strong beam loading. Finally, dose-weighted time-of-flight calculations indicate that the broadband energy distribution still allows our LPA source to irradiate at an effective pulse dose rate of $10^{12}$ Gy/s and average dose rates of tens of Gy/min. These results establish a robust LPA operating regime relevant to the development of preclinical VHEE irradiation sources for medical physics and radiation biology.

\end{abstract}


\maketitle
\thispagestyle{arxivfirst}
\begingroup
\renewcommand{\thefootnote}{*}
\footnotetext[1]{Present Affiliation: CERN, 1211 Geneva 23, Switzerland.}
\endgroup

\section{Introduction}
Very-high-energy electron (VHEE) beams (>50~MeV) are gaining significant interest in the radiotherapy community\cite{desrosiers150250MeVElectron2000, panainoVeryHighEnergyElectron2025} as a promising alternative to photons largely due to their ability to produce a better volume conformity\cite{parshenkovShapingRadiotherapy2025} and deliver doses deeper than conventional intermediate-energy electrons (CIEE) beams\cite{hogstromReviewElectronBeam2006}. Laser-driven plasma accelerators (LPAs) \cite{tajimaLaserElectronAccelerator1979} are promising alternatives to RF-based accelerators for production of high-energy electron beams in compact setups \cite{malkaPrinciplesApplicationsCompact2008}. Their ability to generate ultrashort electron bunches (femtosecond - picosecond duration \cite{lundhFewFemtosecondFew2011}) makes them particularly attractive for applications requiring high instantaneous dose rates like studies of ultra-high dose-rate (UHDR) electron irradiation \cite{esplenPhysicsBiologyUltrahigh2020,rongaBackFutureVery2021}. For fixed spectral and transport parameters, the dose per shot/pulse is directly proportional to the electron beam charge, and the average dose rate is proportional to both the charge per shot and the repetition rate. Therefore, increasing the charge per shot while maintaining a stable spectral distribution and pointing is crucial for the development of LPA-based VHEE sources for preclinical irradiation studies, especially when the available repetition rate is limited. Typically, bunch charges of the order of 100~pC or greater within the VHEE window are desirable\cite{smallVHEEFacilitiesEurope2025} to achieve sufficient average dose rates and reduce irradiation times at the repetition rates of current VHEE-capable LPAs.
In such high-charge operating regimes, variations in laser parameters (energy, focus-depth, pointing) and plasma density can result in substantial shot-to-shot fluctuations in electron charge and energy spectrum for LPAs\cite{jalasBayesianOptimizationLaserPlasma2021}. The injected bunch itself can strongly modify the plasma wake through beam loading \cite{wilksBeamLoadingPlasma1987,tzoufrasBeamLoadingNonlinear2008}. It is comparatively straightforward to alter the capacity of charge injection in the ionization-induced injection scheme \cite{pakInjectionTrappingTunnelIonized2010,mcguffeyIonizationInducedTrapping2010} by changing the concentration of a high-Z dopant. If operating in the nonlinear blowout regime \cite{luNonlinearTheoryRelativistic2006}, the injected bunches causing beam-loading can modify the accelerating field in the bubble. This can perturb the trajectories of the sheath electrons forming the plasma wake structures. In strongly loaded or overloaded wakes, these modifications can substantially alter the structure and evolution of subsequent wakefield structures     as well. 
Previous experimental and numerical studies have reported related dynamics in high-charge acceleration regimes \cite{guillaumePhysicsFullyLoadedLaserPlasma2015, gotzfriedPhysicsHighChargeElectron2020} where the primary electron beam can even become the dominant driver of the wakefields.

In this work, we investigate a regime aimed at maximizing the electron charge contained within the 50~MeV - 200~MeV range while maintaining a reproducible spectral distribution during irradiation of biological targets. Rather than treating beam loading and wake perturbation as a purely detrimental effect, we investigate the possibility of accelerating multiple injected populations to increase the usable charge within a selected energy interval. Since our LPA-based irradiation setup is intended for direct irradiation without energy-selective transport, a broad spectrum can be useful if it contains substantial charge in the VHEE range. Its suitability for a given target also depends on the resulting penetration and dose distribution. The source, in these conditions, was successfully used to investigate irradiation of biological targets \cite{giaccagliaMultiscaleRadiobiologicalAssessment2026,giaccagliaLaserdrivenVHEEPulsed2025}. Using a single-stage LPA with a gas-jet plasma source, we identify an operating regime producing up to 559~pC of total accelerated charge using $\sim$2.4~J laser-pulse energy. The LPA-setup produced a dominant energy component around 50~MeV and a high-energy plateau extending to approximately 150~MeV. Over 30 consecutive shots, the most stable operating point exhibits a pointing jitter of approximately 0.7~mrad, together with low shot-to-shot charge variation. Particle-in-cell simulations are used to investigate the acceleration dynamics underlying the observed spectral structure. The simulations reproduce the formation of two distinct electron populations associated with different stages of the wake evolution and show that substantial acceleration can persist despite strong modification of the plasma wake. This source-performance and acceleration-physics characterization establishes a baseline for future preclinical VHEE studies.

\section{Experimental and Numerical Methods}
\label{sec:exp-setup}
\subsection{Laser-plasma accelerator and gas target}
The Salle Jaune facility at LOA provided the laser source for the experiments. It delivers linearly polarized pulses with a central wavelength of 795 nm and pulse duration (FWHM) of 29.6~fs. 
\begin{figure*}[!t]
    \centering
    \begin{subfigure}[t]{0.47\textwidth}
        \vspace{0pt}\centering
        \includegraphics[width=\linewidth,height=0.21\textheight,keepaspectratio]{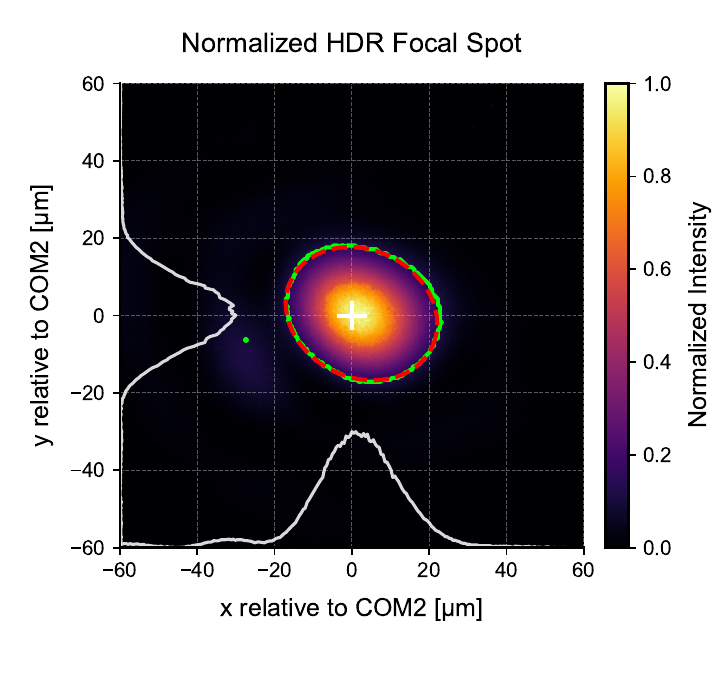}
        \caption{Focal spot measurement of the laser beam at the target plane. The ellipse fit on the focal spot is shown in red and measured semi-major and semi-minor axes are 20.6 $\upmu$m and 17 $\upmu$m respectively.}
        \label{fig:focal-spot}
        \vspace{0.8em}
        \includegraphics[width=\linewidth,height=0.16\textheight,keepaspectratio]{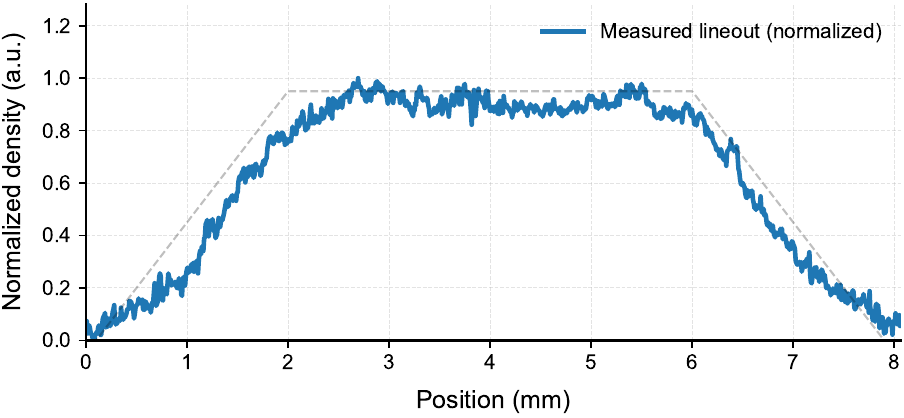}
        \caption{Reconstructed on-axis gas density profile of the nozzle 3~mm above the base. The approximation of the density profile used in the simulations is also shown (dashed line).}
        \label{fig:gas-density-profile}
    \end{subfigure}\hfill
    \begin{subfigure}[t]{0.47\textwidth}
        \vspace{0pt}\centering
        \includegraphics[width=\linewidth,height=0.41\textheight,keepaspectratio]{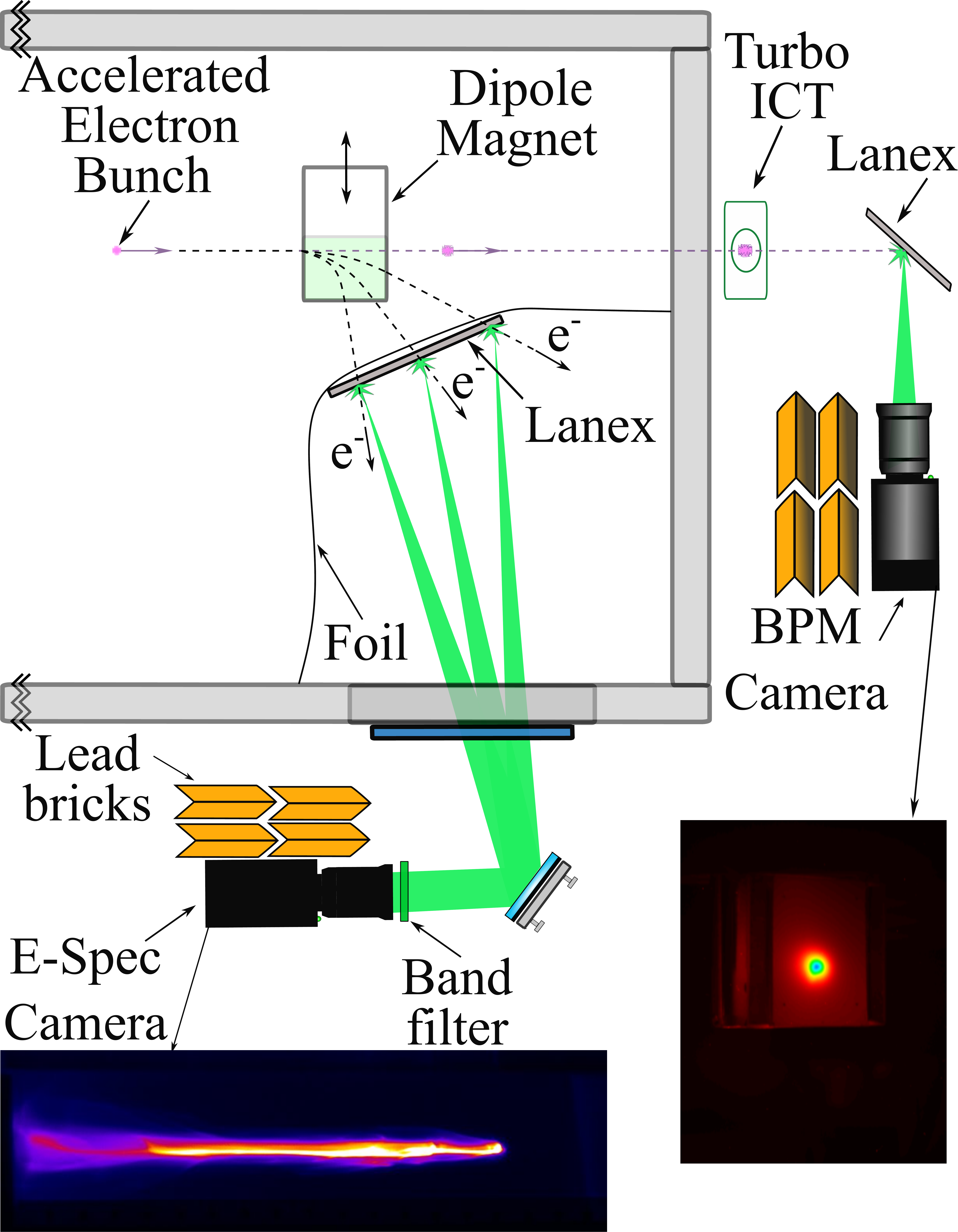}
        \caption{Schematic of the diagnostic setup used for characterizing the electron beams. The irradiation targets were placed 16~cm downstream of the exit flange of the vacuum chamber and were just upstream of the dedicated BPM.}
        \label{fig:diags-active}
    \end{subfigure}
    \caption{Laser focal spot, gas density profile and diagnostic setup used for the high-charge VHEE campaign.}
    \label{fig:experimental-setup}
\end{figure*}
The laser beam was focused in an f/20 setup using a spherical mirror (in conjunction with an upstream deformable mirror) to generate a focal-spot of size $\sim$37 $\upmu$m ($1/e^2$) (see Fig.\ref{fig:focal-spot}). With about 53\% of energy contained in the central lobe of the focal spot, peak intensities achieved were on the order of $a_0 \cong$ 1.9. The gas jet used as the plasma source had a rectangular slit. Mixtures of $N_2$ in Helium (1\%, 2\% and 5\%) were used to enable ionization injection. The general density profile from the nozzles was characterized separately using Mach-Zehnder interferometry (see Fig.\ref{fig:gas-density-profile}). A phase shift measurement via a femtosecond probe beam of the density profile, in-situ, gave values of the plateau density in the lower end of the order of $10^{18}$ cm$^{-3}$. 

\subsection{Electron-beam diagnostics}
\label{main:sec:diagnostics}
In order to characterize the electron beams produced by the LPA, three main diagnostics (see Fig. \ref{fig:diags-active}\footnotemark) were set up: an electron spectrometer (E-Spec), a beam profile monitor (BPM), and an integrating current transformer (ICT). The E-Spec was positioned 85 mm downstream from the gas jet and utilized a retractable 1~T dipole magnet to disperse the electrons based on their energy. A Lanex scintillating screen coupled with a 14-bit Andor camera captured the spatially resolved energy distribution of the electron beam. Upon retracting the magnet, the electron beam is let through a 1.5~mm-thick Aluminum exit flange downstream of the plasma source. A Turbo-ICT \cite{nakamuraPicocoulombChargeMeasured2016} was placed for non-destructive charge measurements and downstream of the ICT, at a distance of 800 mm from the source, a BPM consisting of a Lanex screen and a 14-bit Hamamatsu camera was installed to capture the transverse profile of the electron beam.
\footnotetext{Adapted using Dr. S. Smartsev's original figure}
\subsection{Metrics used for stability analysis}
\label{main:sec:metrics}
For each dataset, the mean spectrum $\bar{S}(E)$ and its shot-to-shot standard deviation $\sigma_S(E)$ were evaluated over the selected energy interval $W$. The spectral variation was quantified by an RMS coefficient of variation,
\begin{equation}
\label{main:eq:metric-1}
    \mathrm{RMS-CV}_{W} =\left[\frac{\int_W \sigma_S^2(E)\mathrm{d}E}{\int_W \bar{S}^2(E)\mathrm{d}E}\right]^{1/2}.
\end{equation}
In this work, $W$ was the VHEE region, $E>50~\mathrm{MeV}$. This metric weights the relative spectral fluctuation by the signal strength, so low-signal spectral tails do not dominate the result. Instead of measuring just the shot-to-shot variation of total charge, it measures the reproducibility of the spectral shape and amplitude over the chosen energy range. Because $\mathrm{RMS-CV}_W$ integrates variance bin-by-bin across the energy axis, small shot-to-shot shifts in peak energy or high-frequency spectral modulations introduce local variance that directly drives up the metric, even when the overall charge yield remains constant. 
Datasets with $\mathrm{RMS-CV}_{E>50~\mathrm{MeV}}<0.30$ were classified as spectrally stable for the present comparison. This is an operational selection criterion: it corresponds to an RMS spectral fluctuation below $30\%$, and is not intended as a universal stability threshold. The VHEE content was additionally defined as the fractional spectral charge above $50~\mathrm{MeV}$,
\begin{equation}
\label{main:eq:metric-2}
    f_{\mathrm{VHEE}} =\frac{\int_{50~\mathrm{MeV}}^{E_{\max}} \bar{S}(E) \mathrm{d}E}{\int_{E_{\min}}^{E_{\max}} \bar{S}(E) \mathrm{d}E}.
\end{equation}
The relative shot-to-shot fluctuation of charge was reported as its coefficient of variation. For the BPM, the transverse centroid was obtained from an ellipse fit to the thresholded beam image. The pointing fluctuations in the two transverse directions were used to calculate a single scalar pointing jitter, 
\begin{equation}
\label{main:eq:metric-3}
    \sigma_{\mathrm{pointing}}=\sqrt{\sigma_{x_{\mathrm{c}}}^{2}+\sigma_{y_{\mathrm{c}}}^{2}}.
\end{equation}

\section{High-Charge VHEE Source Performance}
\label{main:sec:performance}

Optimal conditions used for direct irradiations showed ICT measurements of total bunch charges up to (559$\pm$56) pC for the best case dataset, while dataset-averaged charges/shot were typically 380~pC - 450~pC.
\begin{figure}[!htbp]
       \centering
       \includegraphics[width=\linewidth]{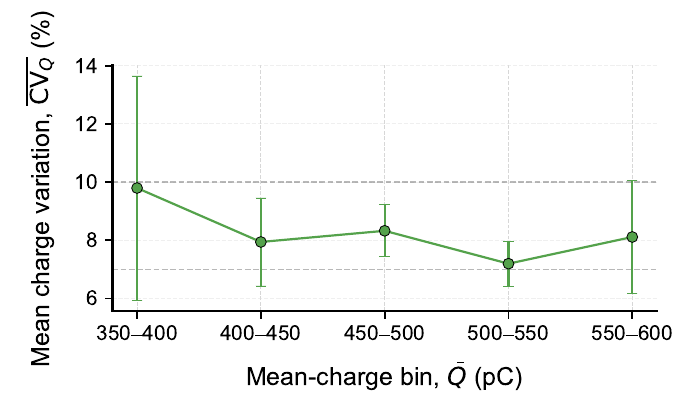}
       \caption{Mean charge variation of datasets ($\overline{\mathrm{CV}_Q}$) as a function of the mean charge yield ($\bar{Q}$). 
       Error bars represent the standard deviation of $\mathrm{CV}_Q$ for the datasets comprising each bin.}
       \label{fig:charge-stability}
\end{figure}
When observing inter-dataset variations, Figure \ref{fig:charge-stability} shows that the relative shot-to-shot charge variation ($\overline{\mathrm{CV}_Q}$) did not degrade at high-charge datasets. Instead it remained bounded between 7\% and 10\% for all mean-charge bins above 350 pC. Notably, the lowest variation datasets were recorded in the 500 - 550 pC bin, demonstrating reliable generation of stable, high-charge electron beams. 
\begin{figure}[!htbp]
       \centering
       \includegraphics[width=\linewidth]{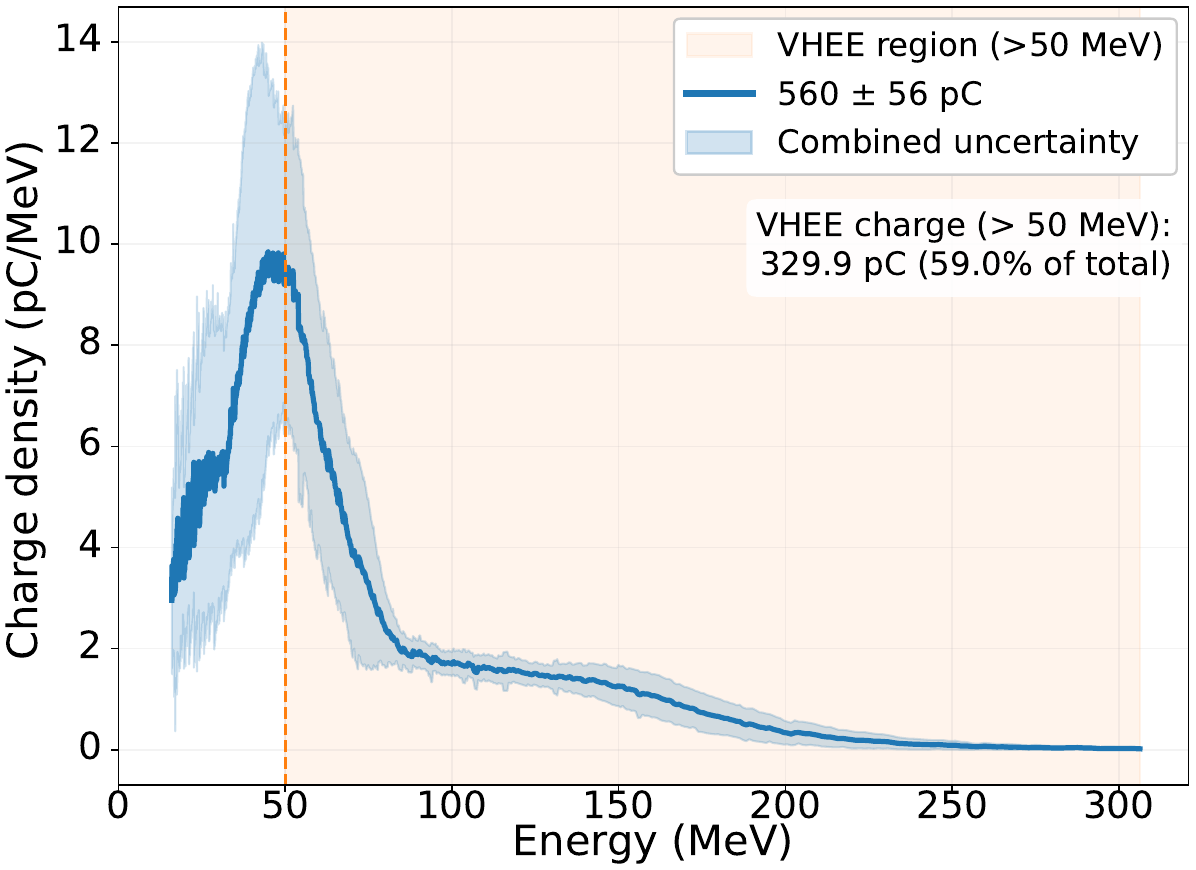}
       \caption{Mean electron-spectra dataset using 30 shots. Prominent peaking at $\sim$50~MeV with a high-energy tail extending up to $\sim$200~MeV.}
       \label{fig:representative-spectra}
\end{figure}
After tuning, the electron beams consistently exhibited a pronounced peak around 40-60 MeV, a high-energy plateau starting from approximately 70-80 MeV extending up to 150 MeV and cutoff slightly below 250~MeV (see Fig. \ref{fig:representative-spectra}). The fraction of the charge in the VHEE range $\geq$50~MeV ranged from 55$\%$ to 70$\%$ for this operating regime while the data across the campaign had the VHEE fraction range from 49$\%$ to 84$\%$. The measured spectra were reproducible across fixed-setting shot sequences, with dataset-averaged RMS-CV values below 30$\%$ over the analysed VHEE interval. The shot-to-shot spectral charge content variations were less prevalent in the high-energy tail compared to the 50~MeV peak region.
\begin{figure}[!htbp]
       \centering
       \includegraphics[width=\linewidth]{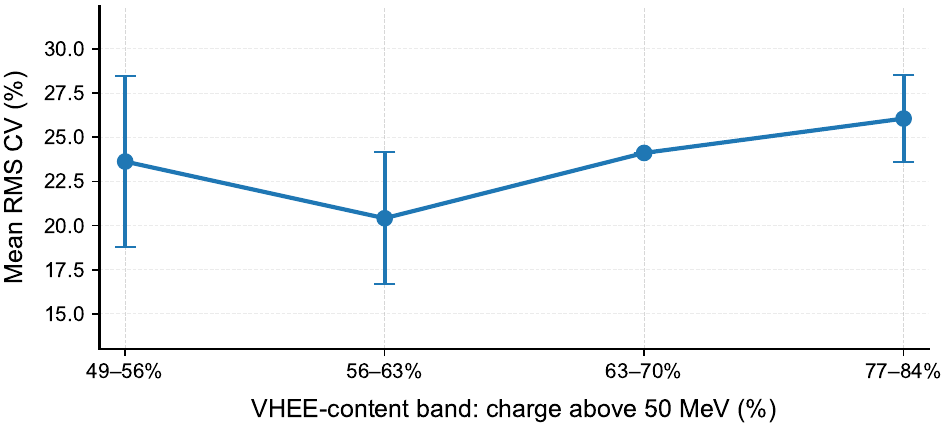}
       \caption{Mean spectral variation (in ${E>50~\mathrm{MeV}}$) measured across different datasets grouped in VHEE-fraction bands. Error bars represent the standard deviations in each bin.\protect\footnotemark}
       \label{fig: spectra_stability}
\end{figure}
\footnotetext{Only 1 dataset was present in the 63-70 $\%$ bin so no error bars are visible.}

Across relevant datasets, it is possible to distinguish stable spectral datasets using the VHEE fraction. A dataset with higher VHEE fraction with lower RMS-CV would be desirable for irradiation applications. Banding our data into VHEE-fraction ranges allows us to see if increasing VHEE content in beams was affecting the shot-to-shot stability in spectral measurements. Across these bands, the RMS-CV$_{E>50~\mathrm{MeV}}$ was found to be below 30$\%$ (see Fig. \ref{fig: spectra_stability}). The minimal spectral variation was observed in the 56\% - 63\% VHEE content band, which demonstrated a mean RMS-CV of approximately 20.5\%. Notably, the most stable individual dataset within this band recorded an $\mathrm{RMS-CV}_{E>50~\mathrm{MeV}}$ of less than 15$\%$ at a 62\% VHEE fraction.

\begin{figure}[!htbp]
       \centering
       \includegraphics[width=\linewidth]{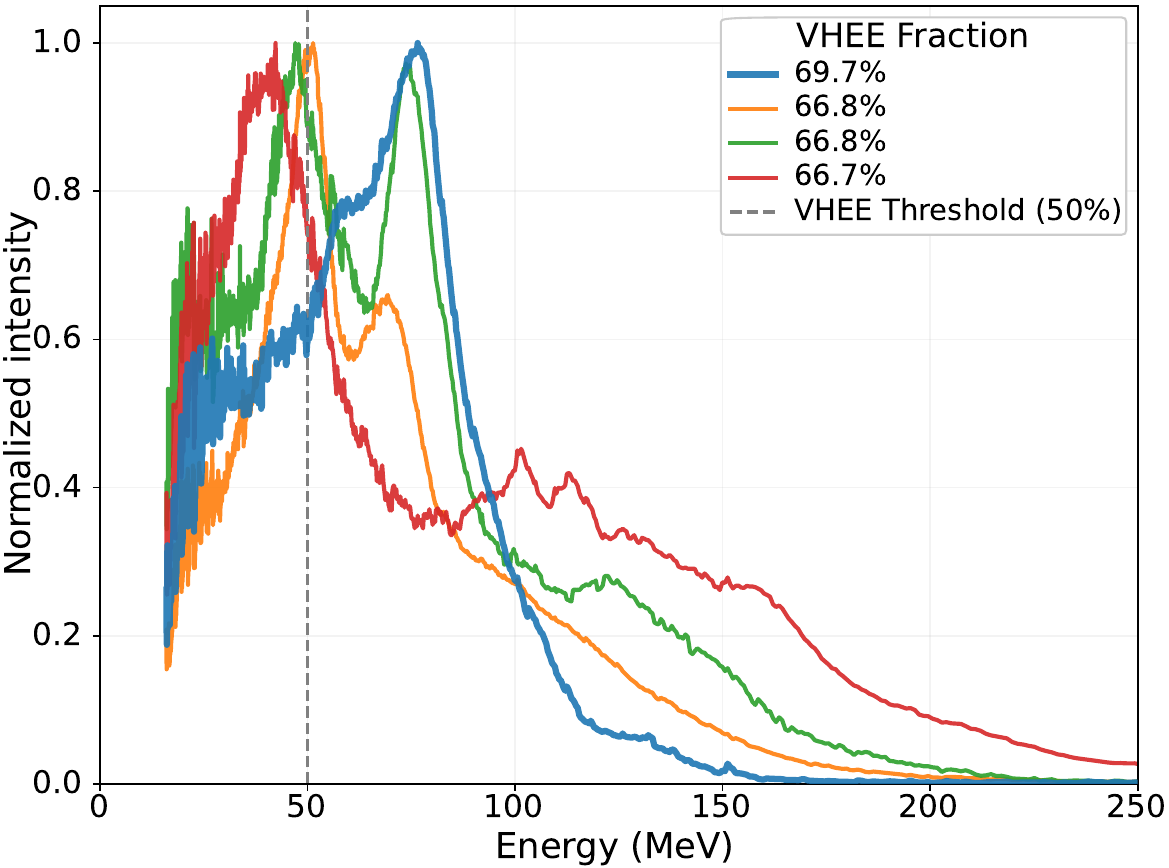}
       \caption{Variations in the mean electron spectra producing different VHEE charge content in datasets.}
       \label{fig:vhee-content-variation}
\end{figure}

Apart from identification of the stable operating points, when parameters like temporal compression, focal position and backing pressure were varied, the spectral features would deviate from the peak and high-energy plateau structure. The high-energy plateau was either enhanced or suppressed (see Fig. \ref{fig:vhee-content-variation}) and the low peak would move to higher energies. Such variation of the spectral shapes observed in the experiments was changing the VHEE content of the beam but was not used for many irradiation experiments as the spectral reproducibility was not guaranteed.

The BPM measurements include a convolution of scattering effects within the exit flange and the original transverse distribution of the beam just after the source. Regardless, these measurements provide realistic variation trends of the relative pointing and beam-profile stability at the irradiation plane. Across the experimental campaign, when measured from the plasma-source, the dataset-averaged pointing jitter ranged from 0.7 mrad to 2.7 mrad. Although the pointing jitter was below 2 mrad for most datasets, beams containing a higher concentration of low-energy electrons were observed to have higher area variations at the irradiation plane. For irradiation applications, we want to make sure that the amount of area in which the dose is deposited is large enough to cover the irradiation target and that the area variations of the beam profile are minimal. 
\begin{figure}[!htbp]  
    \centering
    \includegraphics[width=\linewidth]{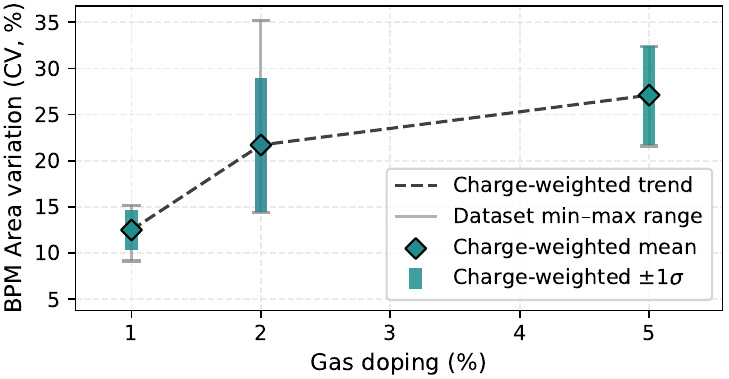}
    \caption{Variation of the beam-profile-area at the irradiation plane as a function of the gas doping percentage. The error bars represent the range of the area variation in datasets for each doping percentage.}
    \label{fig:area-variation}
\end{figure}
The dataset-level coefficients of variation of the BPM area ranged from approximately 9$\%$ to 35$\%$ (shot-to-shot) as seen in Fig.~\ref{fig:area-variation}. As the doping percentage increased, the measured area variation would vary from 12$\%$ (charge-weighted-mean for 1$\%$ doping) to 27$\%$ (charge weighted mean for 5$\%$ doping). Overall, these measurements indicate that the operating regime maintained under 3~mrad pointing stability across the campaign, whereas fluctuations in the transverse beam area represented the larger source of spatial variability at the irradiation plane.
\begin{figure}[!htbp]
    \centering
    \includegraphics[width=0.9\linewidth]{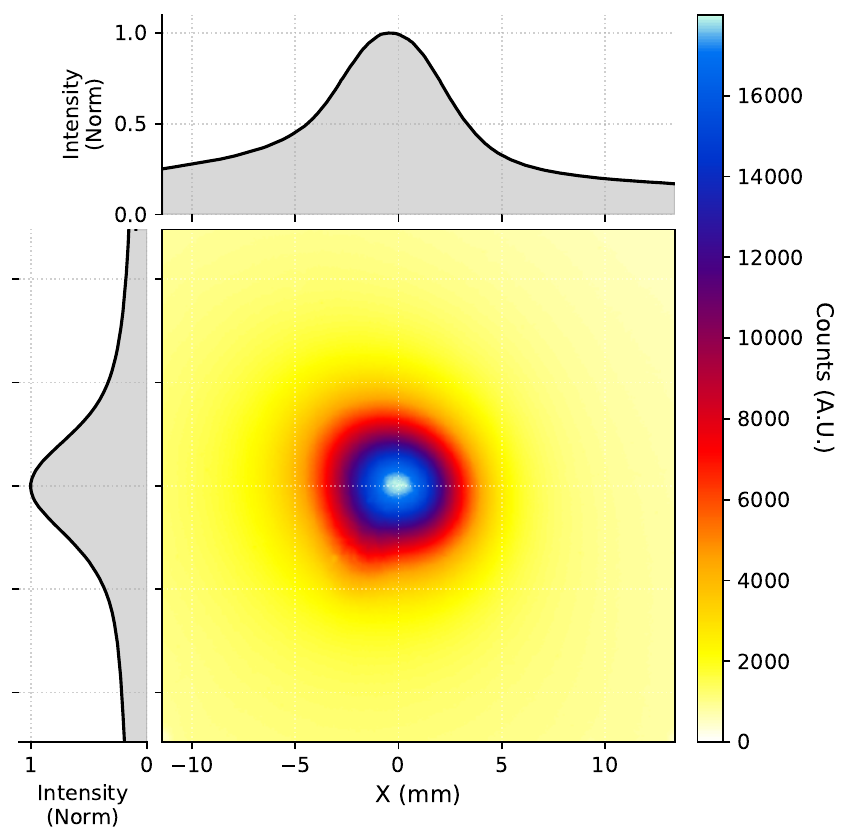}
    \caption{Stability analysis over 30 shots of a representative dataset showing the integrated beam profile measured 800 mm downstream of the source.}
    \label{fig:representative-beam-profile}
\end{figure}
A representative 30-shot dataset shown in Fig.~\ref{fig:representative-beam-profile} illustrates the spatial properties of a comparatively stable operating point. The measured centroid fluctuations were 0.35~mm and 0.48~mm in the horizontal and vertical directions, respectively. For this measurement, the beam-area variations were only 11.9\% w.r.t. the average beam size.

\section{Acceleration dynamics via PIC simulations}
\label{sec:pic-dynamics}
The PIC simulations (which we perform using FBPIC\cite{leheSpectralQuasiCylindricalDispersionFree2016}) allow us to visualize the plasma dynamics that come into play during the acceleration process. Using quasi-cylindrical symmetry to reduce the computational cost, we use the experimental parameters with certain simplifications to perform a Lorentz-boosted simulation. The laser was initialized with a Gaussian temporal profile and a transverse profile imitating the measured focal spot ($\omega_0$ = 18.5~$\upmu m$), with a peak vacuum-propagation $a_0$ of 1.86.
The plasma density profile was reconstructed from the interferometric measurements as described earlier and was simplified into a 2~mm up-ramp, 4~mm plateau and a 2~mm downramp for the simulation (see Supplementary Fig.~\ref{fig:sim-setup}). The simulation was run for a total propagation distance of 8.16 mm, allowing for the observation of the full acceleration process and the resulting electron spectrum at the plasma exit. Separate species were used to model the plasma electrons and the injected electrons of the nitrogen atoms used to observe the ionization-injection.
The laser pulse undergoes significant evolution during propagation and the relative positioning of the nozzle with respect to the focal depth was observed to have a significant impact on the final electron spectrum.
The case with a plasma plateau density of $2 \times 10^{18}~\mathrm{cm}^{-3}$ and a vacuum laser focus at $6000~\upmu\mathrm{m}$ was selected for its qualitative resemblance to the measured peak-and-tail spectral structure.
\begin{figure}[!htbp]
    \centering
    \includegraphics[width=0.93\linewidth]{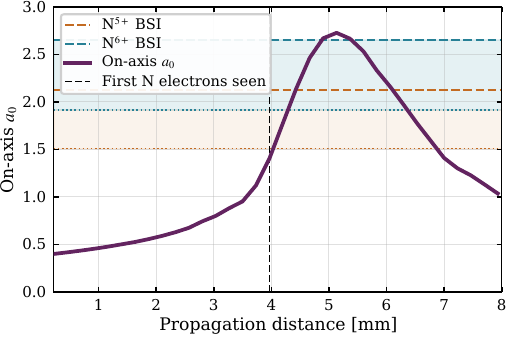}
    \caption{Evolution of $a_0$ along the propagations in the plasma profile. Ionization probability bands are defined in Supplementary S1.}
    \label{fig:a0-evolution}
\end{figure}
A self-focusing effect is observed (as shown in Fig.~\ref{fig:a0-evolution}) as the $a_0$ starts increasing sharply before the center of the plateau ($z\simeq3500~\upmu\mathrm{m}$) and peaks $\simeq 5100~\upmu\mathrm{m}$ into the plasma profile. 
\begin{figure*}[!b]
    \centering
    \includegraphics[width=0.98\textwidth]{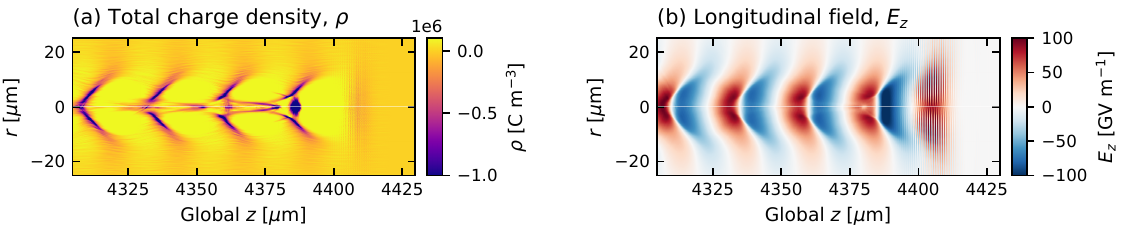}
    \vspace{0.2em}
    \includegraphics[width=0.98\textwidth]{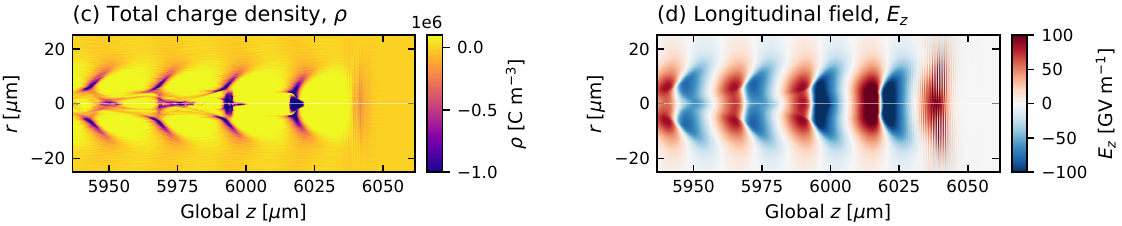}
    \caption{Evolution of the wakefield structure in the simulation. Panels (a, b) show the onset of strong beam loading following ionization injection in the first bubble. Panels (c, d) show the subsequent evolution of the wake and the presence of strong longitudinal electric fields in the second bubble.}
    \label{fig:injection_dynamics_with_e-fields}
\end{figure*}
The pulse-integrated ionization estimates for Nitrogen K-shell electrons are overlaid for comparison. These estimates demonstrate that the field-inferred on-axis laser amplitude coincides with the initial appearance of nitrogen electrons in the simulation. This electron release is governed by an ionization rate calculated using the FBPIC implementation of the ADK model.

Simulations show that injection in the first bubble already disrupts the separation between the first and second bubbles. In particular, the rear sheath of the first bubble extends into the second bubble and no longer closes. This leads to a partial loss of the inter-bubble sheath structure (see Fig.~\ref{fig:injection_dynamics_with_e-fields} (a)). 

Sheath electrons entering the second bubble can be seen following a radially defocused trajectory and a similar partial loss of sheath closure is also observed in the second and third bubbles. Injection of the Nitrogen inner-shell electrons can be tracked as localized variations in the longitudinal electric field, $E_z$ (see Figure~\ref{fig:injection_dynamics_with_e-fields} (b)). These variations coincide with the redistributed charge density associated with the exotic wake-structure formed. Although this macroscopic disruption of the wake perturbs the local fields, a distinct accelerating region is maintained, allowing the first injected population to continue accelerating to higher energies. As the laser propagates deeper in the plasma, $a_0$ peaks, and electron injection into the second bubble also becomes apparent. Particle tracking shows that some of the electrons in the second bubble originate from the same co-moving birth region as electrons in the first bubble. They undergo backward slippage in the bubble disruption before reaching the second accelerating region (see Supplementary Fig.~\ref{fig:slippage_and_births}). Final electron bucket positions therefore do not uniquely specify the bubble in which an electron was born or first trapped. The time-resolved energy-position distributions also show that the injection of the population in the second bucket is already established within the plateau (see also the tracked-particle snapshots in Supplementary Fig.~\ref{fig:tracked-snapshots}). As charge accumulates in the second bunch, the first bunch develops a more conical transverse profile, with its radius increasing toward the rear, as seen in Fig.~\ref{fig:injection_dynamics_with_e-fields} (c). The second bubble is also observed to retain a sufficiently strong accelerating $E_z$ to accelerate the second bunch to high energies (see Fig.~\ref{fig:injection_dynamics_with_e-fields} (d)).
  
\begin{figure*}[!htbp]
    \centering
    \includegraphics[height=5.9cm,keepaspectratio]{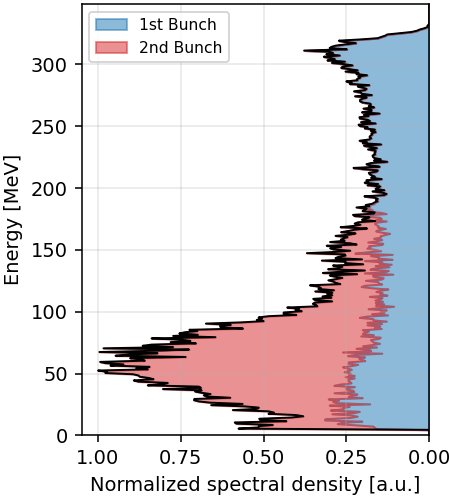}    \hspace{1cm}
    \includegraphics[height=6cm,keepaspectratio]{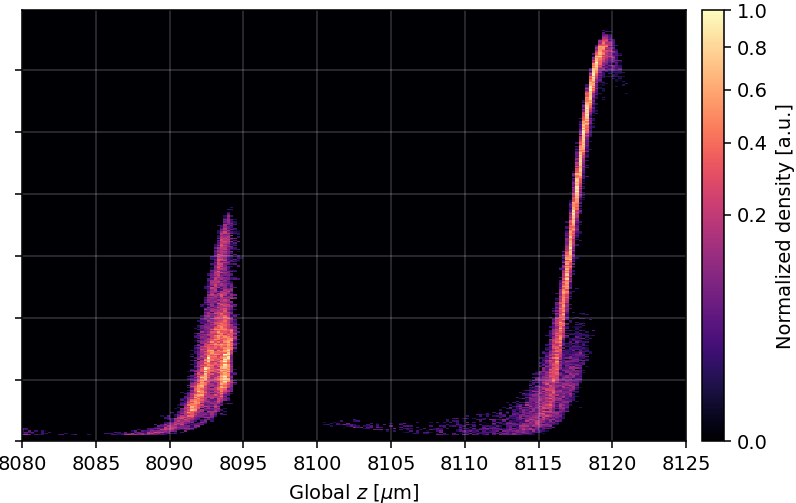}
    \caption{Left: Final simulated electron energy spectrum. Right: Longitudinal spectral distribution of the injected electrons showing two distinct bunch populations. Altered scale to highlight the different energy populations. The blue and red distributions in the left panel identify the individual contributions of the bunches to the final spectrum.}
    \label{fig: sim-final_spectra}
\end{figure*}
 
The final electron spectra observed at the end of the simulation are shown in Fig.~\ref{fig: sim-final_spectra} (left). The simulated spectra show a peak around 50 - 70~MeV along with a high energy tail after the 100~MeV mark qualitatively resembling the experimental peak-and-tail structure.
The population-resolved spectrum in Fig.~\ref{fig: sim-final_spectra} shows that the second bunch makes the dominant contribution to the peak around 50~MeV-70~MeV, while the first bunch supplies the higher-energy component. The corresponding energy - position distribution shows that the two populations remain longitudinally separated at the plasma exit despite overlapping in energy. Their superposition can therefore produce a continuous peak-and-tail spectrum without requiring the bunches to merge longitudinally. This decomposition provides a physical interpretation compatible with the experimentally observed combination of a dominant mid-energy component and a higher-energy tail. The comparison concerns the spectral structure; the two populations and their separate contributions have not been resolved experimentally.

\section{Temporal profile at target plane and effective dose rate}
\label{main:sec:temporal}

\begin{figure}[!htbp]
    \centering
    \includegraphics[width=0.95\linewidth]{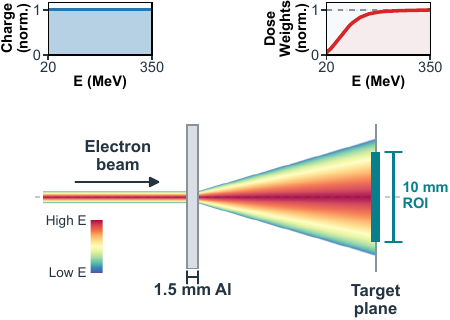}
    \caption{Schematic illustration of energy-dependent ROI-dose weighting. The blue curve represents an illustrative flat incident charge spectrum and the red curve represents the corresponding relative ROI dose contribution.}
    \label{fig:scattering-response}
\end{figure}

For a polychromatic electron beam, different energy electrons with the same charge do not deposit equal dose at the irradiation plane ROI, nor do they arrive at the same time. To estimate the temporal profile of dose deposition at the irradiation plane, 691.5~mm from the source, it is necessary to account for both: the energy-dependent dose contribution and the temporal broadening caused by the large spectral width during propagation. GEANT4-based Monte Carlo simulations~\cite{agostinelliGeant4aSimulationToolkit2003} were used to calculate the energy-dependent dose response in the target plane ROI. Specifically for transport through the aluminium flange (Fig.~\ref{fig:scattering-response}). 
The energy-dependent response $R(E_i)$ gives the dose deposited in the target plane ROI per unit incident charge at kinetic energy $E_i$, in Gy/pC. This response includes the effect of transport through the simulated flange-geometry on the ROI dose. Each population element, representing an incident charge $Q_i$, is therefore assigned a dose contribution $w_i = Q_i R(E_i)$. The temporal distribution of the polychromatic electron bunch must be reconstructed at the target plane rather than inferred directly from the bunch duration at the end of the PIC simulations. In the simulation-informed model, the simulated $(z,E)$ population is reweighted to reproduce the measured energy spectrum while retaining the simulated longitudinal distribution at each energy. For comparison, a second model assumes a uniform longitudinal charge distribution over 30~$\upmu$m (approximately 100~fs), combined with the same measured energy spectrum. 
Using either of the two approaches, the corresponding target-plane arrival time $t_i$ can be estimated using ballistic propagation. 

\begin{figure}[!htbp]
    \centering
    \includegraphics[width=\linewidth]{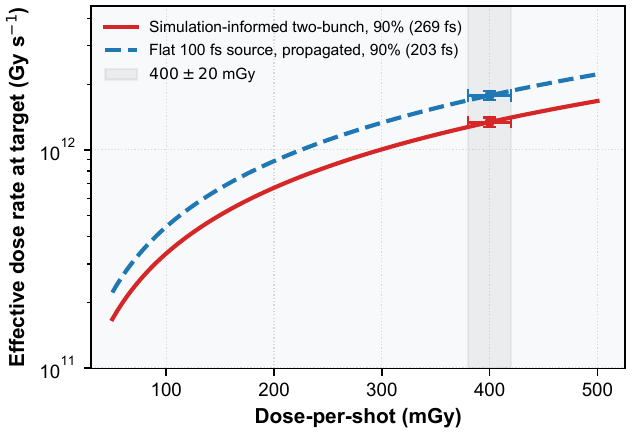}
    \caption{Effective pulse dose rate at the target plane as a function of dose per shot, comparing the simulation-informed two-bunch model with a uniform 100-fs source model.}
    \label{fig:dose-rate}
\end{figure}

Combining the estimated arrival times $t_i$ with the assigned dose contrinutions $w_i$ gives the cumulative delivered-dose fraction,
\begin{equation*}
F(<t)=\frac{\sum_{t_i<t} w_i}{\sum_i w_i}.
\end{equation*}
The independently measured dose per shot, $D_{\rm shot}$, converts this cumulative fraction to an absolute delivered dose:
\begin{equation*}
D(<t)=D_{\rm shot}F(<t).
\end{equation*}
The 90\% dose-delivery time, $t_{90}$, is defined by $F(<t_{90})=0.90$, with time measured from the earliest ballistic arrival in the modelled population. 
\begin{figure}[!h]
    \centering
    \includegraphics[width=\linewidth]{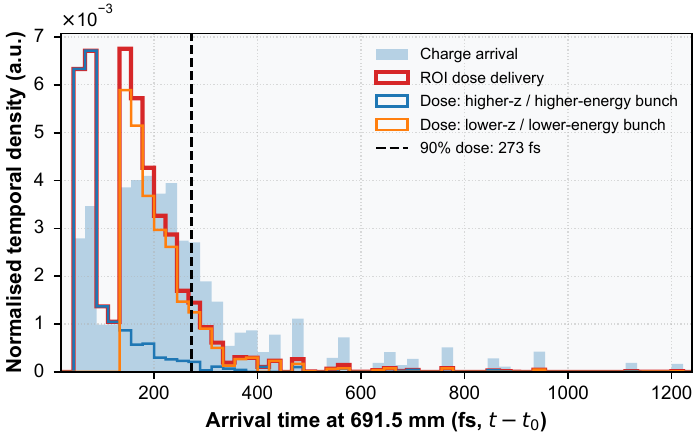}
    \caption{Target-plane arrival distribution showing the dose-weighted contributions of the two simulated electron populations. $t_0$ is the earliest arrival time.}
    \label{fig:arrival-time}
\end{figure}
Applying this to the simulation-informed two-bunch distribution yields a 90\% dose-delivery duration of $t_{90}=269$~fs (see Fig.~\ref{fig:arrival-time}). The dose-weighted arrival distribution clearly shows that the higher-energy primary bunch arrives first, while the secondary bunch elongates the overall delivery window. We define the effective pulse dose rate over this delivery interval as
$$
\dot{D}_{\rm eff,90}=\frac{0.90D_{\rm shot}}{t_{90}}.
$$
Using our independently estimated target-plane dose-per-shot from irradiation measurements, $D_{\rm shot}\sim 0.400\pm0.020$~Gy, we obtain:
$$
\dot{D}_{\rm eff,90} = \frac{0.90\times0.400\ {\rm Gy}}{269\ {\rm fs}} = (1.34\pm0.07)\times10^{12}\ {\rm Gy\,s^{-1}}.
$$
Upon doing the same calculations for a flat-longitudinal-profile $30~\upmu m$ bunch, an effective dose-delivery duration of $t_{90}=203$~fs is obtained and an effective dose rate of $(1.77\pm0.09)\times10^{12}\ {\rm Gy\,s^{-1}}$ (see Fig.~\ref{fig:dose-rate}). 
The estimated $\dot{D}_{\rm eff,90}$ is not a microscopic peak dose rate, but rather a robust, macroscopic pulse-scale metric. By using the 90\% convention, we avoid disproportionate sensitivity to weak, late-arriving electron tails while capturing the vast majority of the biologically relevant dose. 

\section{Discussion and outlook}
For pre-clinical radiobiology research involving targets of limited depth, the 50~MeV threshold is a practical lower bound for VHEE irradiation. In this context, an LPA source can focus on delivering higher doses with stability and reproducibility rather than on producing electron beams with a narrow energy spectrum. The high-charge, broadband spectra observed in this work allow us to deliver more dose within a useful energy interval.

Our LPA source was operated over several days of irradiation experiments spanning three weeks. In some studies, irradiations were carried out on three separate days, so it was important to identify operating points that produced a comparable electron spectrum across sessions. This ensured that changes in the biological response could be interpreted without spectral variation adding another degree of freedom to the dose-rate study. The usage of this source in radiobiological studies of cells, zebrafish, lung slices, and mice, reported in peer-reviewed medical journals, also demonstrates its practical utility for irradiation research \cite{giaccagliaMultiscaleRadiobiologicalAssessment2026,giaccagliaLaserdrivenVHEEPulsed2025}.

In our setup, the exit-flange serves as a vacuum-to-air interface and is designed to block the laser while adding minimal scattering to the VHEE component of the beam. The scattering effect increases the beam size at the irradiation plane and prevents the minimal pointing jitter from causing the beam to miss the target. Further optimization of the extraction and transport geometry could improve the dose-per-shot.

For such beams, using only the electron-bunch duration at the plasma exit can overestimate the effective instantaneous dose rate. The TOF calculations are therefore necessary to estimate a realistic dose-delivery time at the target plane. The difference in values of estimated temporal profiles is a direct result of the assumed longitudinal profile at source. But in both the assumed cases, the calculations show that the majority of the single pulse dose is delivered in less than 300~fs at the chosen irradiation plane. It is important to note that the quantity $\dot{D}_{\rm eff,90}$ should be interpreted as a pulse-scale dose-rate.

The present PIC interpretation of our LPA source provides a mechanism for continued acceleration in successive wake periods during substantial wake restructuring. The comparatively reduced energy gain of the second bucket of electrons may result from a shorter acceleration history, beam loading or changes in the evolving wakefields. The present diagnostics do not allow this contribution to be perfectly isolated. Whether this acceleration mechanism also reduces sensitivity to shot-to-shot input fluctuations remains to be established. The observed spectral reproducibility may reflect a repeatable balance between ionization injection and beam loading. Increased loading could partially compensate an increase in the unloaded wake amplitude, while redistribution between successive wake periods could reduce fluctuations in the combined useful charge. The comparatively low centroid fluctuations, despite larger transverse-profile variations, are consistent with approximately symmetric beam evolution about a stable propagation axis. These possibilities require simulations with perturbed input conditions to be tested.

Although the experiments do not demonstrate independent control of the two populations, the observed sensitivity of the spectral shape to focal position, gas pressure and laser compression suggests that the relative contribution of the lower-energy component and the high-energy tail can be modified through the accelerator conditions. Understanding whether this corresponds to control over the relative injection into the two wake periods would be an important next step. If this relationship can be established, optimization of the source could be carried out in terms of the charge contained above a chosen energy threshold rather than only the position or width of a spectral peak.

In conclusion, the source performance characterized here supports the use of a reproducible broadband LPA regime for direct VHEE irradiation in preclinical applications. The simulations suggest that injection into successive wake periods contributes to the useful charge, while the dose calculations connect the resulting spectrum to delivery at the target. Next steps would be to determine how each electron population and its fluctuations contribute to target dose and delivery time. This would guide source optimization toward reproducible irradiation conditions for specified targets.

\begin{center}\textbf{Acknowledgement}\end{center}
The authors thank Igor Andriyash, Cédric Thaury and Jérôme Faure for the helpful discussions around the PIC simulation. The authors acknowledge the use of OpenAI Codex to assist with LaTeX formatting and manuscript copy-editing. All suggestions were reviewed by the authors, who take full responsibility for the final content.

\begin{center}\textbf{Funding}\end{center}

AF acknowledges Foundation of Ecole Polytechnique (grant BIO-RAPIDE HC), Ile-de-France Region (grant SESAME/IDRA), and ANR contract FemtoDose. AF and CG acknowledge CNRS contract MITI/BioRapide.

\begin{center}\textbf{Competing interests}\end{center}
The authors declare no competing interests.
\begin{center}\textbf{Author contributions}\end{center}

CV and CG set up and ran the experiment with general supervision from AF. JG, JPG, AT, PR provided the experimental area and laser system support. CV conceived the paper, performed the post-experiment analysis and PIC modelling, and wrote the paper with corrections from AF. All authors contributed to the discussion of the results.

\begin{center}\textbf{Data availability}\end{center}

The data supporting the findings of this study are available from the corresponding author upon reasonable request.

\section*{Supplementary material}

\setcounter{figure}{0}
\renewcommand{\thefigure}{S\arabic{figure}}
\setcounter{equation}{0}
\renewcommand{\theequation}{S\arabic{equation}}

\subsection*{S1. Simulation setup and ionization estimates}
\label{supp:sec:setup}
\begin{figure}[!htbp]
    \centering
    \includegraphics[width=\linewidth]{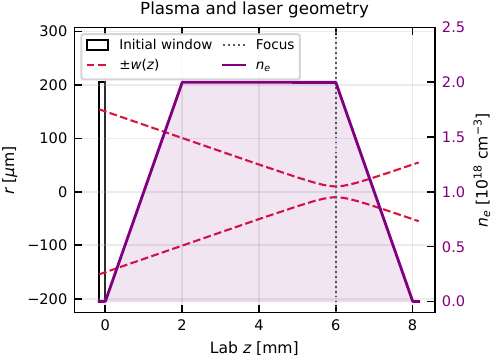}
    \caption{Representative PIC geometry: prescribed plasma density, initial moving window and laser focusing geometry at the end of the plateau, $z=6~\mathrm{mm}$.}
    \label{fig:sim-setup}
\end{figure}

Figure~\ref{fig:sim-setup} shows the PIC simulation setup that uses a radially symmteric laser focal distribution and a simplified plasma density profile, not considering the additional variations in laser pointing, spectral phase and gas-target conditions. 
In Fig.~\ref{fig:a0-evolution}, the orange and blue shaded bands provide illustrative ionization-amplitude ranges for the nitrogen K-shell transitions.
The lower, dotted boundaries denote a chosen pulse-integrated ADK ionization probability of $15\%$, evaluated for a Gaussian pulse with an intensity FWHM of $29.6~\mathrm{fs}$ at $\lambda_0=795~\mathrm{nm}$. The upper, dashed lines denote approximate instantaneous BSI levels. 

\subsection*{S2. Particle histories and final-population labels}
\label{supp:sec:histories}
\begin{figure}[!htbp]
    \centering
    \includegraphics[width=\linewidth]{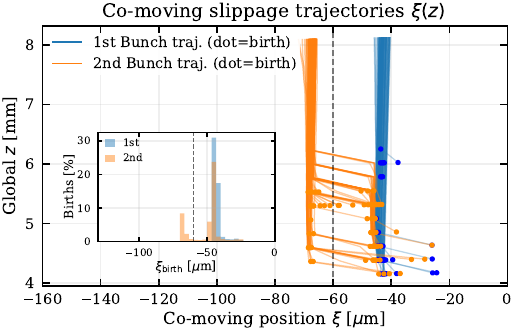}
    \caption{Co-moving electron trajectories, labelled by final bucket membership: blue for the first group and orange for the second. Dots mark electron birth by ionization. The inset compares the birth-coordinate distributions. The vertical axis of the main plot is the global propagation distance. The dashed line at $\xi=-60~\upmu\mathrm{m}$ is the displayed grouping reference, rather than a tracked wake boundary.}
    \label{fig:slippage_and_births}
\end{figure}

The trajectories in Fig.~\ref{fig:slippage_and_births} complement the final spectrum and energy - position distribution in Fig.~\ref{fig: sim-final_spectra}. They relate the final population labels to electron `birth'/ionization of an electron. The inset shows the corresponding birth-coordinate distributions. 

Final-population labels are retained throughout each history. Particle-point densities illustrate motion, rather than charge-weighted yields.
\FloatBarrier

\onecolumn
\subsection*{S3. Selected stages of the tracked-particle evolution}
\label{supp:sec:snapshots}
\begin{figure}[!htbp]
    \centering
    \begin{subfigure}{\linewidth}
        \centering
        \includegraphics[width=0.93\linewidth,height=0.33\textheight,keepaspectratio]{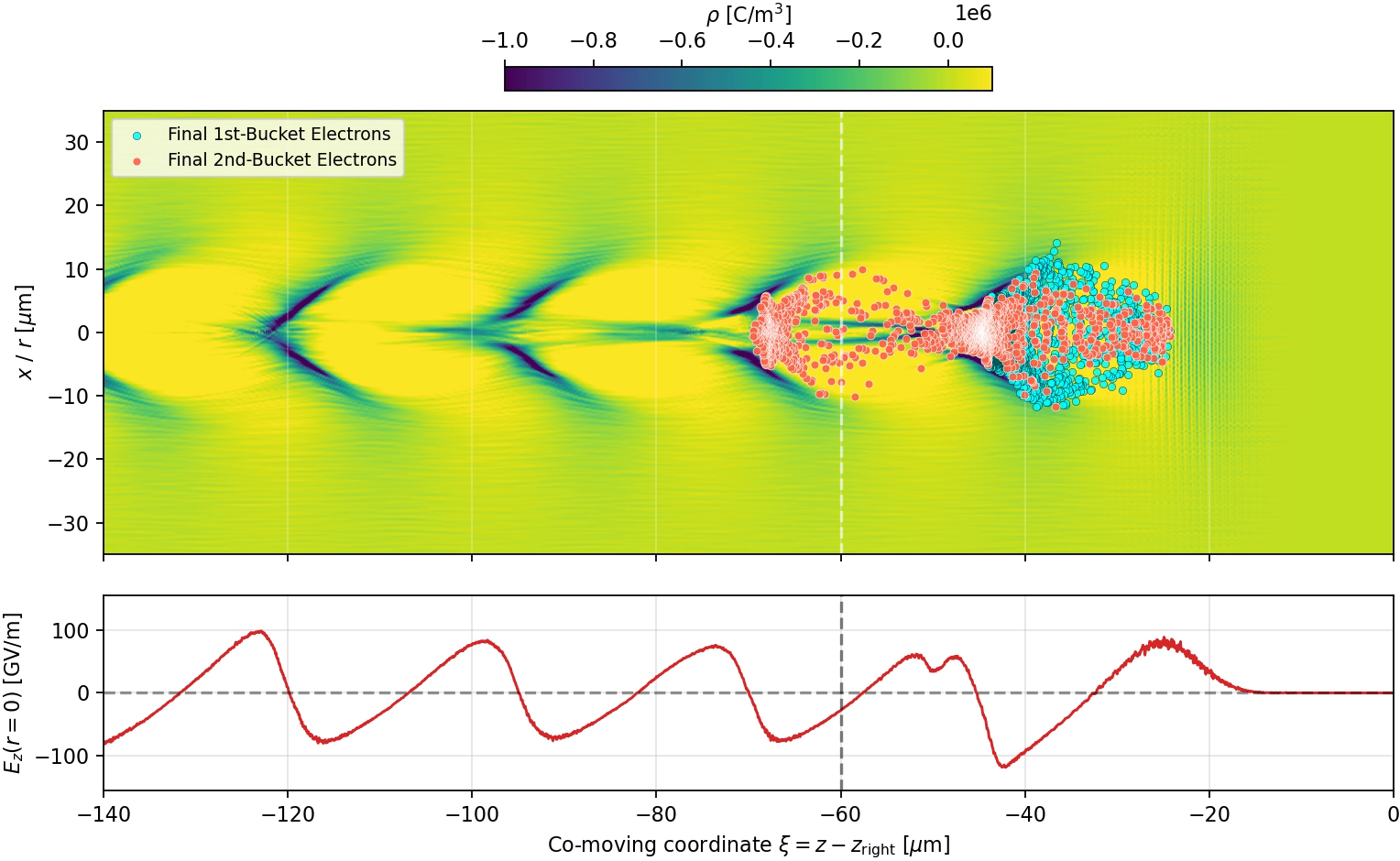}
        \caption{$t_{\rm lab}=14.78~\mathrm{ps}$; $z_{\rm right}=4.43~\mathrm{mm}$.}
    \end{subfigure}\par\medskip
    \begin{subfigure}{\linewidth}
        \centering
        \includegraphics[width=0.93\linewidth,height=0.33\textheight,keepaspectratio]{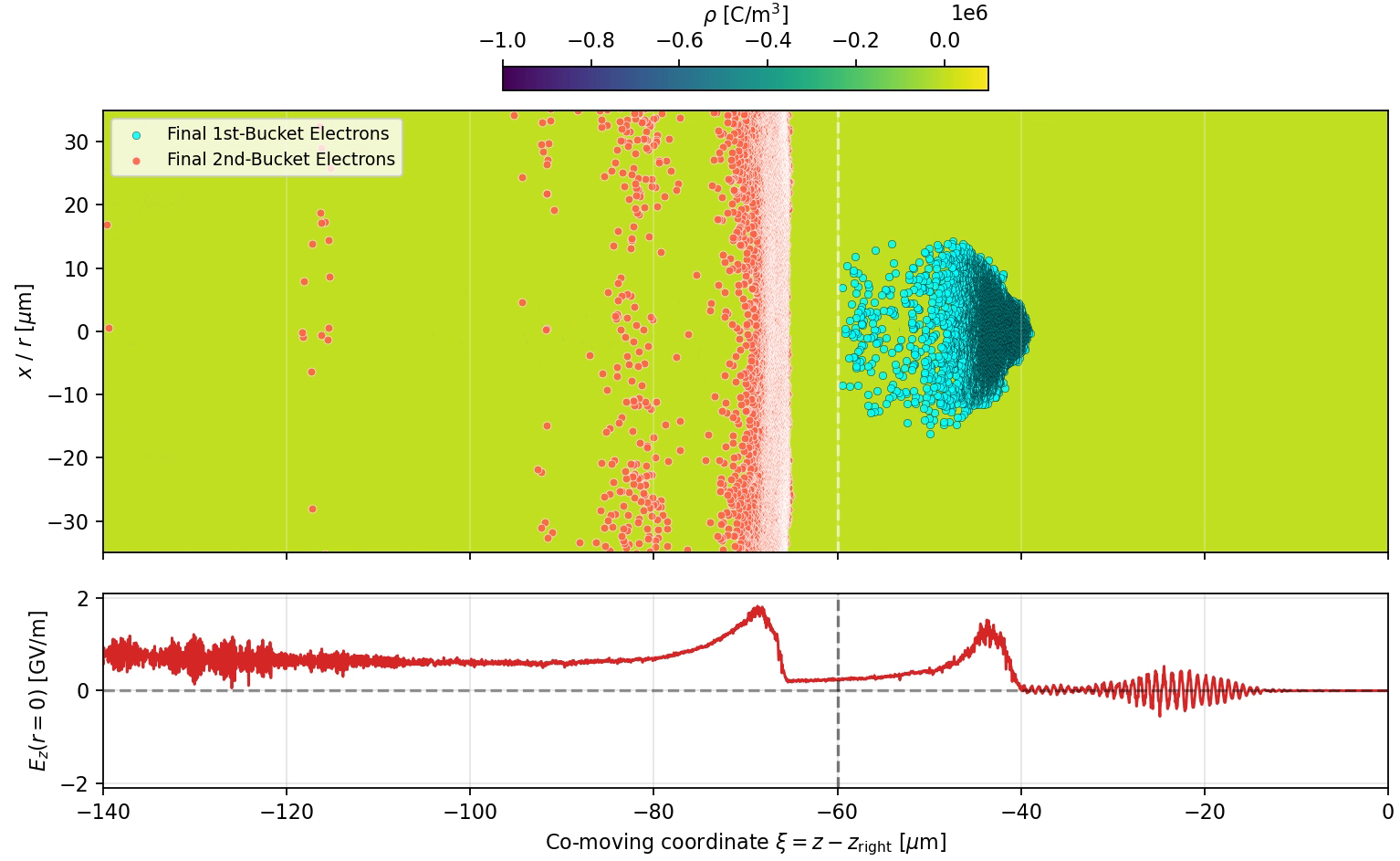}
        \caption{$t_{\rm lab}=27.23~\mathrm{ps}$; $z_{\rm right}=8.16~\mathrm{mm}$.}
    \end{subfigure}
    \caption{Selected stages of the tracked-particle evolution. Each panel shows the charge density with selected particle positions (upper plot) and the on-axis longitudinal electric field (lower plot). Cyan and red retain the final first- and second-bucket labels at both stages. The longitudinal electric-field scales differ between (a) and (b). Particle markers illustrate spatial histories and are not a charge-density scale.}
    \label{fig:tracked-snapshots}
\end{figure}

Figure~\ref{fig:tracked-snapshots} shows the selected particles overlaid on the charge density, together with the on-axis longitudinal electric field, at $z_{\rm right}=4.43$ and $8.16~\mathrm{mm}$. The co-moving coordinate is $\xi=z-z_{\rm right}$, where $z_{\rm right}$ denotes the right-hand reference position of the displayed window. Cyan and red denote the final first- and second-bucket groups, respectively, corresponding to the blue and orange groups in Fig.~\ref{fig:slippage_and_births}. 

At the earlier stage, particles assigned to the two final groups occupy overlapping longitudinal regions during wake restructuring. At the plasma exit, the selected groups are spatially separated. The field profiles provide the instantaneous wake context, but do not by themselves quantify energy transfer between the laser, the bunches and the plasma.

\twocolumn

\bibliographystyle{unsrt}
\bibliography{paper_citation_correction,all_possible_refs}

\end{document}